# Prominent room temperature valley polarization in WS₂/graphene heterostructures grown by chemical vapor deposition

I. Paradisanos[1,2], K. M. McCreary[3], D. Adinehloo[4], L. Mouchliadis[1], J. T. Robinson[5], Hsun-Jen Chuang[3], A. T. Hanbicki[3], V. Perebeinos[4], B. T. Jonker[3], E. Stratakis[1,6] and G. Kioseoglou[*1,6]

[1] Institute of Electronic Structure and Laser, Foundation for Research and Technology - Hellas, Heraklion, 71110, Crete, Greece

[2] Department of Physics, University of Crete, Heraklion, 71003, Crete, Greece

[3] Materials Science and Technology Division, Naval Research Laboratory, Washington, DC 20375, USA

[4] Department of Electrical Engineering, University at Buffalo, The State University of New York, Buffalo, NY 14260, USA

[5] Electronics Science and Technology Division, Naval Research Laboratory, Washington, DC 20375, USA

[6] Department of Materials Science and Technology, University of Crete, Heraklion, 71003 Crete, Greece

We examine different cases of heterostructures consisting of WS₂ monolayers grown by chemical vapor deposition (CVD) as the optically active material. We show that the degree of valley polarization of WS₂ is considerably influenced by the material type used to form the heterostructure. Our results suggest the interaction between WS₂ and graphene (WS₂/Gr) has a strong effect on the temperature dependent depolarization (i.e. decrease of polarization with increasing temperature), with polarization degrees reaching 24% at room temperature under near-resonant excitation. This contrasts to hBN- encapsulated WS₂, which exhibits a room temperature polarization degree of only 11%. The observed low depolarization rate in WS₂/Gr heterostructure is attributed to the nearly temperature independent scattering rate due to phonons and fast charge and energy transfer processes from WS₂ to graphene. Significant variations in the degree of polarization are also observed at 4K between the different heterostructure configurations. Intervalley hole scattering in the valence band proximity between the K and Γ points of WS₂ is sensitive to the immediate environment, leading to the observed variations.

* Corresponding author: gnk@materials.uoc.gr

The band structure of transition metal dichalcogenides (TMDs) has unique features that makes them ideal candidates for valleytronics, a field where the valley-index is a potential new state variable.[1–6] The low-dimensional hexagonal lattice structure of TMDs, combined with strong orbital hybridization and time-reversal symmetry result in two inequivalent, high symmetry points, K/K´, with coupled spin and valley indices and unique optical selection rules. An imbalance in the carrier population between the K/K´ valleys, referred to as a valley polarization (VP) can therefore be created under excitation with circularly polarized light, enabling the independent initialization and addressing of the valley-index. Research on a more fundamental understanding of all the parameters affecting the VP is ongoing. Some of the parameters that can affect the polarization include excitation energy,[4,7] temperature,[7] electron-hole exchange interactions,[8] and electron-phonon coupling.[4,9] Although there are several reports on the effect of the surrounding background on the electronic properties of TMDs,[10–20] only a few examine the effect of temperature, excitation energy and environment on the valley polarization.[21,22] Further investigation is expected to provide additional insight into the physics of TMDs in different surrounding environments.

In this work, we investigate the T-dependent VP properties of chemical vapor deposition (CVD) synthesized WS$_2$ monolayers and their heterostructures. In particular, T-dependent (4K-300K) polarization phenomena and the related scattering mechanisms of neutral excitons (X$^0$) are examined in near-resonant and off-resonant conditions for WS$_2$ on graphene (WS$_2$/Gr) and hBN/WS$_2$/hBN heterostructures. Because of the absence of charged excitons (trions, X$^-$) in the WS$_2$/Gr system, we only focus on the properties of X$^0$ in order to compare the exact same recombination channel. Interestingly, significant differences are observed in the low-temperature valley-polarization degree, *P(4K)*, as well as in the rate the polarization drops as a function of temperature. In addition, a surprising nearly T-independent VP was observed in WS$_2$/Gr under near-resonant pumping conditions. These differences are investigated through T-dependent electron-phonon coupling and changes in the band-structure between $WS_2/Gr$ and $hBN/WS_2/hBN$. This study offers an insight into fundamental phenomena of 2D TMD monolayers towards the development of future large-scale, fast, low-cost, environmentally sustainable optoelectronic devices.

Here we focus on $WS_2/Gr$ and $hBN/WS_2/hBN$ heterostructures. Comparison with standard $WS_2/SiO_2$ and $WS_2/hBN$ (non-encapsulated) samples are also provided when needed, however, we note that $WS_2/Gr$ and $hBN/WS_2/hBN$ have superior stability and spatial uniformity with respect to $WS_2/SiO_2$ and $WS_2/hBN$. Fig.1a presents optical images of $hBN/WS_2/hBN$, $WS_2/Gr$. A portion of the $WS_2$ in the $hBN/WS_2/hBN$ sample is not encapsulated giving us an access to a $WS_2/hBN$ sample.

The degree of circular polarization is extracted from helicity-resolved photoluminescence (PL) spectra; therefore, it is important to first examine the different emission channels in each system. In Fig.1b we compare the micro-PL spectra of the different cases when excited with a 543nm (2.283 eV) laser source at low temperature (4K). Two distinct features at 2.083eV and 2.046eV are present in $WS_2/SiO_2$. We assign these features to $X^0$ and $X^-$, respectively.[23] In the case of $WS_2/Gr$, there is only one peak (2.042 eV). This feature is due to $X^0$ emission and is red shifted 41meV compared to $WS_2/SiO_2$. The assignment of this as $X^0$ is established with differential-reflectivity measurements (Fig.1c). The spin orbit (SO) energy difference between the B- and A-exciton is preserved, as both are red shifted equally, compared to the $SiO_2$ substrate (supplementary-Fig.S1). The observed shift is attributed to the presence of the underlying graphene which screens the electric field between electrons and holes and leads to a strong reduction in the bandgap and more modest reduction of the exciton binding energy of the monolayer (ML) $WS_2$.[14] The absence of $X^-$, as well a 6-fold suppression of the $X^0$ intensity compared to $WS_2/SiO_2$ (Fig.S2) is a result of nonradiative recombination processes such as photogenerated charge carrier transfer to graphene due to strong interlayer electronic coupling.[24]

In the case of hBN/WS$_2$/hBN, the suppression of disorder effects by the crystalline, wide band-gap hBN substrate leads to a reduction of the excitonic emission linewidths, compared to the WS$_2$/SiO$_2$ and facilitates the identification of additional peaks.[15,19,20] Multiple features in the top panel of Fig.1b include X$^0$, X$^-$ and a biexciton (XX). For the hBN/WS$_2$/hBN system, X$^0$ is centred at 2.065eV, and XX at 2.002eV.[25] For the X$^-$ that is asymmetric up to 80K and cannot be fitted well by a single function, two pseudo-Voigt functions are superimposed with an energy difference of 10meV (Fig.S3a). The lower energy feature corresponds to intravalley trions (X$^-_{intra}$-2.021eV, one-hole and two-electrons in the same valley) while the higher energy to intervalley trions (X$^-_{inter}$-2.031eV, the two electrons originate from different valleys, i.e. K and K′).[26,27] Similar effects are observed in the X$^-$ of WS$_2$/hBN (Fig. S3b), where the positions of the X$^-_{intra}$ (2.041eV) and X$^-_{inter}$ (2.049eV) result in an energy difference of 8meV. The intensity of X$^0$ in hBN/WS$_2$/hBN is also considerably suppressed compared to WS$_2$/hBN. This can be attributed to cavity effects[28], as well as the n-type character of WS$_2$. It has been shown that exposure to air can deplete excess electrons in WS$_2$ and enhance the X$^0$ emission (e.g. physisorbed O$_2$ and H$_2$O).[29,30]

The SO splitting in the valence band plays a crucial role in the VP phenomena. For this reason, reflectance contrast spectra, $\Delta R/R = (R_{substrate} - R_{sample})/R_{substrate}$, are collected at 4K (Fig.1c) to examine possible variations in the energy difference between the A and B ground state excitonic resonances due to changes in the dielectric environment. In all cases, the SO splitting is of the same order (400meV–Fig.S1) with negligible variations indicating similar behavior of the A- and B-excitonic state and ruling out effects that could arise from shifts in the energy splitting at the K-points. More detailed fitting of the reflectance contrast spectra (Fig.S1) as well as Raman measurements (Fig.S4a) and a discussion of strain effects are presented in the supplementary-section D.

To measure the polarization, the system is excited with right-handed circularly polarized light ($\sigma_+$), and then the resultant photoluminescence is analyzed for co-polarized ($\sigma_+$) and cross-polarized/left-handed circularly polarized light ($\sigma_-$). The degree of valley polarization is related to the circularly polarized emission ($P_{circ}$)[1,2] where $P_{circ} = (I_+ - I_-)/(I_+ + I_-)$, and $I_+(I_-)$ is the intensity of the $\sigma_+$ or $\sigma_-$ helicity PL component. In Figs.1d and 1e, the $X^0$ T-dependent polarization degree of hBN/WS$_2$/hBN and WS$_2$/Gr are shown for non- and near-resonant excitation conditions using the 543nm (2.283eV) and 594nm (2.087eV) laser sources, respectively. Typical temperature-dependent $\sigma_+$ and $\sigma_-$ PL spectra are demonstrated in Fig.S3c. Slight variations in $P_{circ}$ across different areas of the same monolayer are observed, likely related to spatial non-uniformities in the electron density of the sample.[31,32] For this reason statistical analysis over several points has been applied and the mean values and their standard deviations are presented.

For non-resonant excitation, $P_{circ}$ of the $X^0$ emission at room temperature is nearly zero for both hBN/WS$_2$/hBN and WS$_2$/Gr (Fig. 1d). Both systems exhibit a monotonic increase in the polarization with decreasing temperature. However, the final polarizations reached at cryogenic temperatures are distinctly different for the two systems; $P_{circ}$=41% for hBN/WS$_2$/hBN and $P_{circ}$=19% for WS$_2$/Gr at 4K. This will be addressed later.

Excitation energy is a significant parameter when studying VP phenomena since phonon assisted intra and intervalley scattering can be enabled as a function of excess energy, $\Delta E = E_{excitation} - E_{emission}$.[4,7] To evaluate this effect, we have performed experiments using a 594nm laser source, which is very close to the excitonic emission of WS$_2$, to reduce the excess energy introduced into our systems. The T-dependent $P_{circ}$ under near-resonant conditions, is plotted for both systems in

Fig.1e. As might be expected, overall the values of $P_{circ}$ acquired at these temperatures are higher compared to those measured with the non-resonant 543nm excitation because of the suppresion of the excess energy which would otherwise introduce additional intervalley scattering. More surprisingly, however, $P_{circ}$ of $X^0$ from the WS$_2$/Gr system is nearly temperature independent exhibiting a valley polarization close to 30% in excellent agreement with previous reports using Mueller polarimetry analysis and in MoS$_2$.[21,22] This results in a rather robust $P_{circ}$ of 24% in WS$_2$/Gr even at 300K. Note that experimental issues, related to a strong contribution from Raman scattering make the corresponding value of the polarization degree unreliable below 100K. This point is discussed in more detail in the supplementary (section E, Fig.S5). The behavior of the T-dependent polarization degree in WS$_2$/Gr under near-resonant and non-resonant excitation conditions, is further verified by a second sample presented in FigS3d.

For steady-state conditions the measured polarization can be rationalized by[4]:

$$P_{circ} = \frac{P_0}{1+2\frac{\tau_r}{\tau_v}} \ , (1)$$

where $\tau_v$, is the overall intervalley scattering time (valley relaxation time) and $\tau_r$, is the effective exciton lifetime (including both radiative and non-radiative). Here $P_0$ is the initial polarization of the system which is considered to be equal to 1. We can model the T-dependence of $P_{circ}$ by considering screening effects due to carrier doping as has been done elsewhere.[33,34] This analysis together with the fitting of the experimental points of Fig.1d is presented in the supplementary (section-F, Fig.F1). Yet, it should be noted that the excitation energy strongly affects the T-dependence and this model is not sufficient to fit the data under near-resonant conditions. In addition, we should comment that taking into account similar doping densities in WS$_2$ when it is encapsulated in hBN or on top of graphene is not a realistic scenario since graphene affects the electron-density of WS$_2$. On the other hand, if we change the electron density (determining T-dependence of the Thomas-Fermi wave-vector), we do not get a reliable fit (discussed in the supplemental-section F).

At this point, we'll focus on the observed low-depolarization gradient of the $WS_2$/Gr compared to hBN/$WS_2$/hBN, under non-resonant excitation (543nm) and -more importantly- near-resonance (594nm) shown in Fig.1d and 1e, respectively. Several mechanisms could be responsible for the temperature insensitive valley-polarization of $WS_2$/Gr which we discuss in the following paragraphs.

First, the presence of the linear dispersion of graphene close to the K-point of $WS_2$, gives rise to rapid charge and energy transfer processes via near-field interactions.[35,36,37] The drastically shortened exciton lifetime (~ps scale) in TMD/graphene heterostructures has been demonstrated by both cryogenic[37] and room temperature TRPL experiments[36]. The fingerprint of these processes includes a 14.7meV broadening in the excitonic absorption due to the initiation of a new decay channel for $WS_2$ excitons (Fig.1c).[35] The fitting analysis of the reflectivity spectra is included in section A of the supplementary material (Fig.S1). As a result, following the generation of e-h pairs in the K-valley, a portion of the population that would otherwise be subjected to intervalley scattering processes will now rapidly recombine and transfer the photoexcited energy to graphene non-radiatively, preserving the polarization.[21,22] However, a critical parameter we should take into consideration is the T-dependent intervalley scattering rate under resonant and non-resonant excitation. Considering the electron-phonon coupling in the two heterostructures, we have calculated the scattering rates a) from the bottom of the conduction band at the K-point to all available states near the K´-point (Fig.2a) and b ) at excitation energies large enough to scatter an optical phonon in hBN (100meV above the exciton resonance) which has a stronger electron-phonon scattering (Fig.2b). Note that the exciton scattering rates would be roughly twice as large since both an electron and a hole forming an exciton can scatter a phonon. Calculations details can be found in the supplementary (section-H, Fig.S7). It is clear from Figs.2a,b that under either resonant or non-resonant excitation the scattering rates are almost T-independent because of the large phonon energy. Thus, the main contribution to the observed VP should come from changes in the radiative lifetime. The radiative and the total lifetime of the exciton evolve with temperature. The former is known to increase with temperature[38] while the transfer time to graphene is in the picosecond range and presumably much more weakly T-dependent than the radiative lifetime. Therefore, the fast transfer mechanism is expected to be the dominant factor in the total lifetime in $WS_2$/Gr even at elevated temperatures, in contrast to hBN/$WS_2$/hBN. As a result, the excitonic lifetime in $WS_2$/Gr will be quite weakly T-dependent and the $P_{circ}$ is not expected to vary significantly. It should be noted that a strong correlation between the exciton radiative rate and the thickness of hBN encapsulation was recently revealed as a consequence of the Purcell effect, therefore, the VP degree can be further tailored.[39] It has also been established that the electron-hole exchange-interaction plays a notable role in the polarization in TMD monolayers.[8,40] However, this demanding approach should be treated independently in future studies by taking into account proximity effects from graphene.

Second, in Fig.2c, the PL intensity of $X^0$ after 543nm excitation is plotted as a function of temperature and normalized at 4K. From this figure significant variations are observed between the two cases. While there is a clear enhancement of $X^0$ starting at ~40K in hBN/$WS_2$/hBN, for $WS_2$/Gr the intensity remains almost constant until ~170K. These intensity variations are a consequence of thermal dissociation of $X^-$ and thermally assisted dark-to-bright state transitions. Thermal dissociation of $X^-$ will increase the population of $X^0$ in hBN/$WS_2$/hBN and this pathway could further introduce intervalley scattering. On the contrary, in $WS_2$/Gr, where $X^-$ is not formed, this effect is minimized. Note that transitions from the dark to bright state are also expected to enhance the $X^0$ PL intensity at elevated temperatures[41], surprisingly though in the WS$_2$/Gr this effect is negligible, possibly because of energy transfer from the dark state to graphene (Fig.2c).

A final note regarding the thermally stable valley-polarization of WS$_2$/Gr is related to the observed T-dependent $X^0$ emission energy. In Fig.2d, we compare the measured $X^0$ energy between hBN/WS$_2$/hBN and WS$_2$/Gr, normalized at 4K. Interestingly, the T-dependent band-gap renormalization in WS$_2$/Gr is partially suppressed, resulting in ~20meV energy difference at 300K compared to hBN/WS$_2$/hBN. Therefore, the T-dependent excess energy, has a weaker contribution to the depolarization rate in WS$_2$/Gr which further assists the robust room temperature valley-polarization. To investigate the origin of the reduced thermal effect to the band-gap renormalization in WS$_2$/Gr we have performed additional calculations (Fig.S8). The T-dependent bandgap-renormalization due to the intraband polar phonons appears much weaker than what is observed in the experiment. Two reasons could be responsible for this discrepancy: a) only Fröhlich coupling to phonons and intraband scattering were considered in our calculations, whereas there are also other phonons and interband scattering (virtual valence-to-conduction transitions), which we could not take into account and b) the T-dependence of the screening could also modify the energy of the excitons.

We now move to another significant observation, that is the 2-fold enhancement of the hBN/WS$_2$/hBN valley-polarization compared to WS$_2$/Gr at 4K under non-resonant excitation (Fig.1d). This is a non-trivial result considering the shorter recombination lifetime of WS$_2$/Gr.[36,42] To understand this, we first consider the emission and absorption energy of $X^0$. In the WS$_2$/Gr, $X^0$ is red-shifted by >20meV compared to hBN/WS$_2$/hBN, an effect we attribute to dielectric screening. Therefore, under the same excitation energy, WS$_2$/Gr will always have additional excess-energy in the system that can contribute to intervalley-scattering relaxation mechanisms. While excess energy could be part of the answer, this explanation may be insufficient to explain the full difference between $P_{circ}$ = 41% (hBN/WS$_2$/hBN) and $P_{circ}$ = 19% (WS$_2$/Gr) at 4K. Changes in the band-structure of WS$_2$ in the proximity of hBN or graphene will also affect hole scattering mechanisms during relaxation. It has been proposed that spin degeneracy in the Γ-valley can enable incoherent two-step transitions and therefore holes can relax from K to K´ through scattering via the Γ-valley.[43] We have performed band-structure calculations (supplementary-Section G, Fig.S6) and we find there is a considerable shift in the valence band at the Γ-point in WS$_2$/graphene compared to the corresponding band-structure at hBN/WS$_2$/hBN (Fig.3a,3b). For hBN/WS$_2$/hBN, the energy difference between the valence bands at the K and Γ-points of the Brillouin zone is 590meV, whereas in WS$_2$/Gr it is only 261meV. The relative proximity of the Γ-band to the top of the valence band at K/K´ could affect the low T-polarization, $P_{circ}(4K)$, when the structure is excited under non-resonant conditions (2.283eV) and should be considered during intervalley scattering. To model the observed differences in the low-temperature circular polarization, we introduce a simplified model that takes into account the photo-generated holes scattered to Γ-valence bands (Fig.3c). The Γ-valley holes are then re-scattered back to the K/K´ valleys (to reduce their energy) on a time scale $\tau*$ during the electron relaxation process. Therefore, in the presence of Γ-valence bands in close proximity to K and K´, the rate equations for the carrier populations, $N$ and $N´$ in the K/K´ valleys, respectively are modified as following:

$$\frac{dN}{dt} = g - \frac{N}{\tau_r} - \frac{N}{\tau_V} + \frac{N'}{\tau_V} + \frac{\alpha \cdot N}{\tau^*} - \frac{N}{\tau^*} \, , (2)$$

$$\frac{dN'}{dt} = g' - \frac{N'}{\tau_r} - \frac{N'}{\tau_V} + \frac{N}{\tau_V} + \frac{\alpha \cdot N}{\tau^*}, (3)$$

where $g$ and $g´$ are the generation rates of the K/K´-valleys and considering that $\alpha$ is a percentage of holes at Γ with respect to K: $N_\Gamma = \alpha \cdot N$.

In steady state, $\frac{dN}{dt} = 0$ and $\frac{dN'}{dt} = 0$. Assuming $g = 1$ and $g' = 0$,

$$\frac{N'}{N} = \frac{\frac{1}{\tau_r} + \frac{\alpha}{\tau^*}}{\frac{1}{\tau_r} + \frac{1}{\tau_v}}, \quad (4)$$

Therefore, the circular polarization, considering the proximity of the valence bands at K and $\Gamma$, will be:

$$P_\Gamma = \frac{N - N'}{N + N'} = \left( \frac{1}{1 + \frac{2\tau_r}{\tau_v} + \frac{\alpha\tau_r}{\tau^*}} \right) - \frac{\frac{\alpha\tau_r}{\tau^*}}{1 + \frac{2\tau_r}{\tau_v} + \frac{\alpha\tau_r}{\tau^*}} , (5)$$

where it is apparent that, as the energy of $\Gamma$ approaches K/K´, there will be a reduction of the polarization. Hole scattering should be suppressed under near-resonant excitation, since the relaxation of the hole through the $\Gamma$ valence band is not energetically favorable.

In conslussion, we have demonstrated temperature dependent valley polarization measurements in CVD-grown a) ML WS$_2$ encapsulated in hBN and b) ML WS$_2$ on top of graphene excited with two different photon energies (i.e. off-resonant and near-resonant excitation). The results reveal a surprisingly low depolarization rate as a function of temperature in WS$_2$/Gr that is attributed to a) rapid charge and energy transfer processes of the scattered excitons in the K´-valley of WS$_2$ to graphene via near-field interactions, b) absence of thermal dissociation of trions and thermally assisted dark-to-bright state transitions in WS$_2$/Gr and c) partial suppression of the temperature-dependent band-gap renormalization in WS$_2$/Gr. We find that under either resonant or non-resonant excitation the scattering rates are almost Tindependent because of the large phonon energy. Thus, the main contribution to the observed valley depolarization are likely to come from the changes in the total lifetime. Notable differences between hBN/WS$_2$/hBN and WS$_2$/Gr are also presented in $P_{circ}(4K)$ under non-resonant excitation that are attributed to changes in the energy position of the $\Gamma$-valley affecting the activation of hole-scattering channels.

## SUPPLEMENTARY MATERIAL

See supplementary material for further details.

## ACKNOWLEDGEMENTS

The authors thank Gang Wang for the useful discussions. IP acknowledges the financial support of the Stavros Niarchos Foundation within the framework of the project ARCHERS (Advancing Young Researchers Human Capital in Cutting Edge Technologies in the Preservation of Cultural Heritage and the Tackling of Societal Challenges). ES and GK acknowledge financial support from Nanoscience Foundries and Fine Analysis (NFFA)–Europe H2020-INFRAIA-2014-2015 (under Grant agreement No 654360) and MouldTex project-H2020-EU.2.1.5.1 (under Grant agreement No 768705). The work at the U.S. Naval Research Laboratory (NRL) was supported by the Office of Naval Research through core programs at NRL.

**FIGURES**

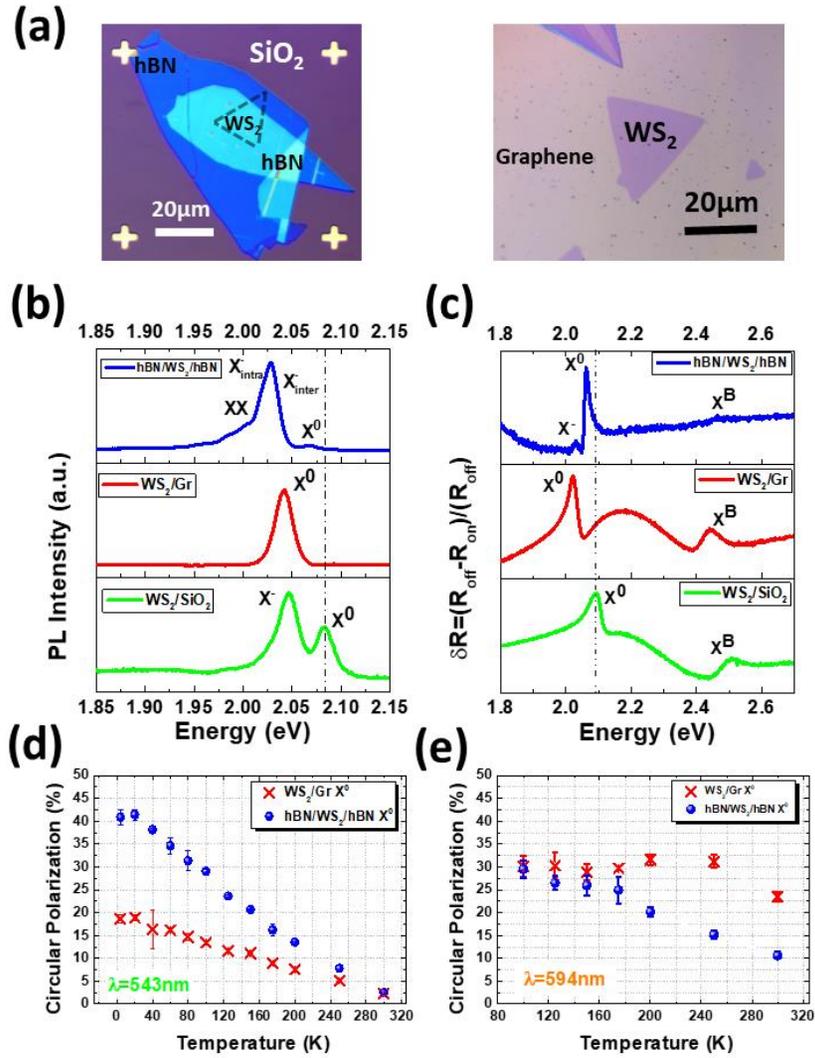

**Figure 1:** (a) Optical images of hBN/WS$_2$/hBN and WS$_2$/Gr. (b) PL spectra at 4 K excited with 543 nm laser and (c) Reflectance contrast spectra at 4K for the three systems measured. In each panel, from bottom to top: WS$_2$/SiO$_2$, WS$_2$/Gr and hBN/WS$_2$/hBN. (d) and (e) T-dependent valley-polarization for X$^0$ of hBN/WS$_2$/hBN and WS$_2$/Gr under 543nm and 594nm excitation, respectively.

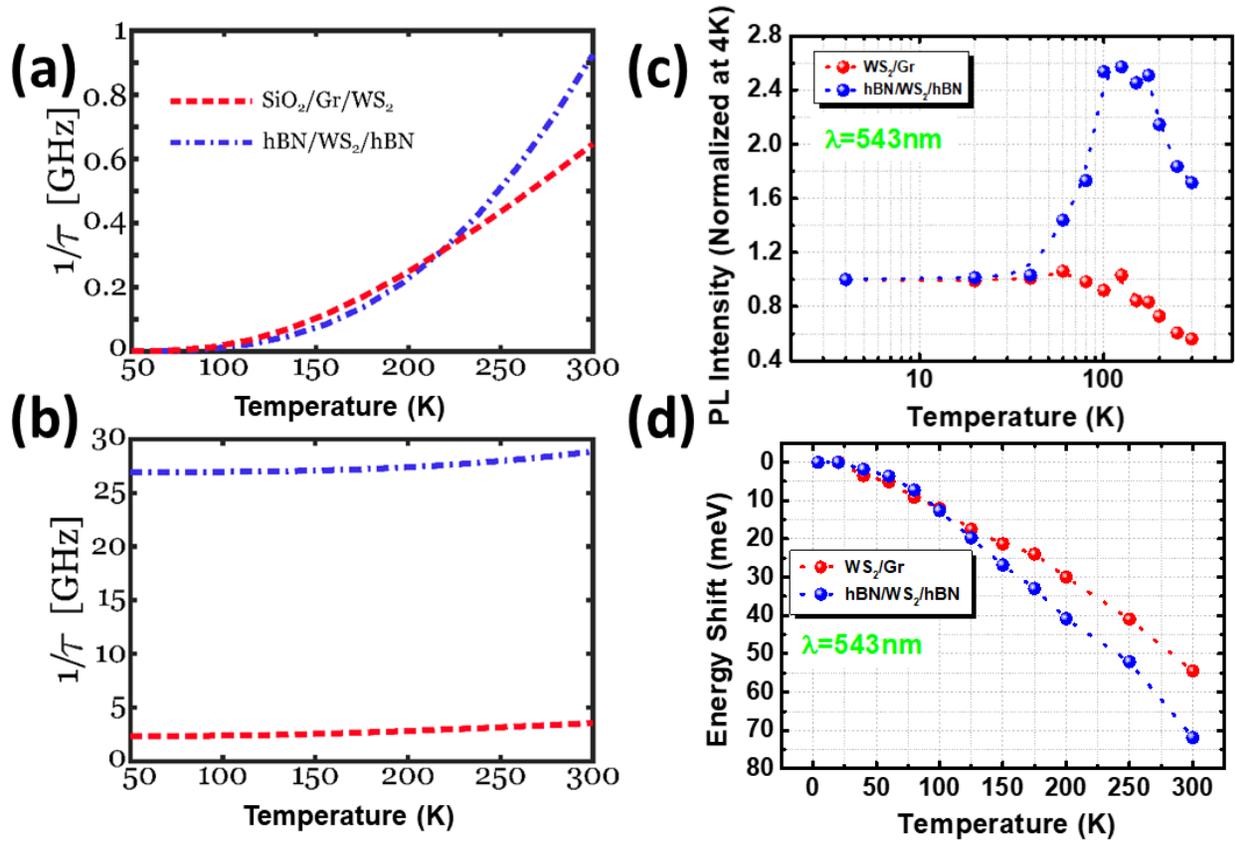

**Figure 2:** Calculated intervalley scattering rates of electrons in WS$_2$ heterostructures under (a) resonant (excess energy, $\boldsymbol{\varepsilon_k = 0\ meV}$) and (b) off-resonant excitation ($\boldsymbol{\varepsilon_k = 200\ meV}$). (c) Normalized (at 4 K) PL intensity of X$^0$ as a function of temperature in hBN/WS$_2$/hBN and WS$_2$/Gr. (d) T-dependent energy shift of X$^0$. The energy shift is normalized and set to 0 at 4K.

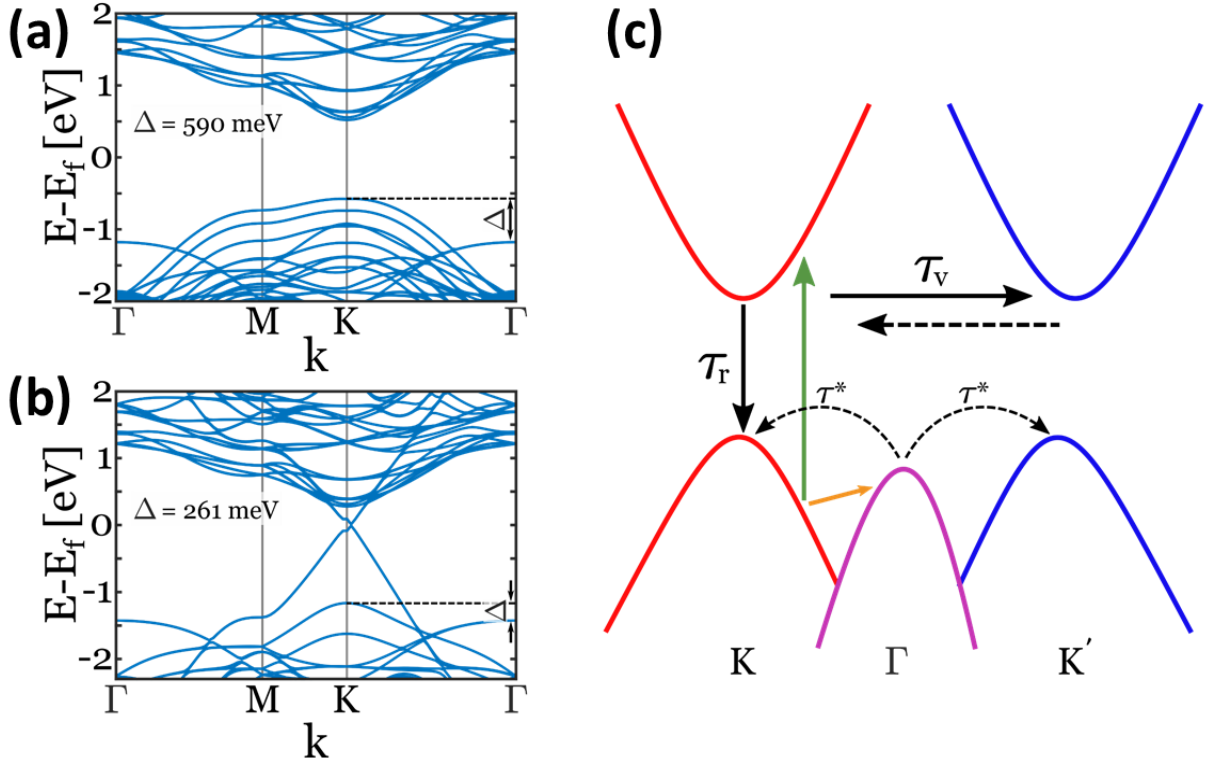

**Figure 3:** Band structure calculations of a) WS$_2$/6layer-hBN and b) WS$_2$/Gr. c) Schematic representation of the hole-phonon intervalley scattering effects in the proximity of Γ/K bands.

# Supplementary Information

## A. *Reflectance contrast spectra*

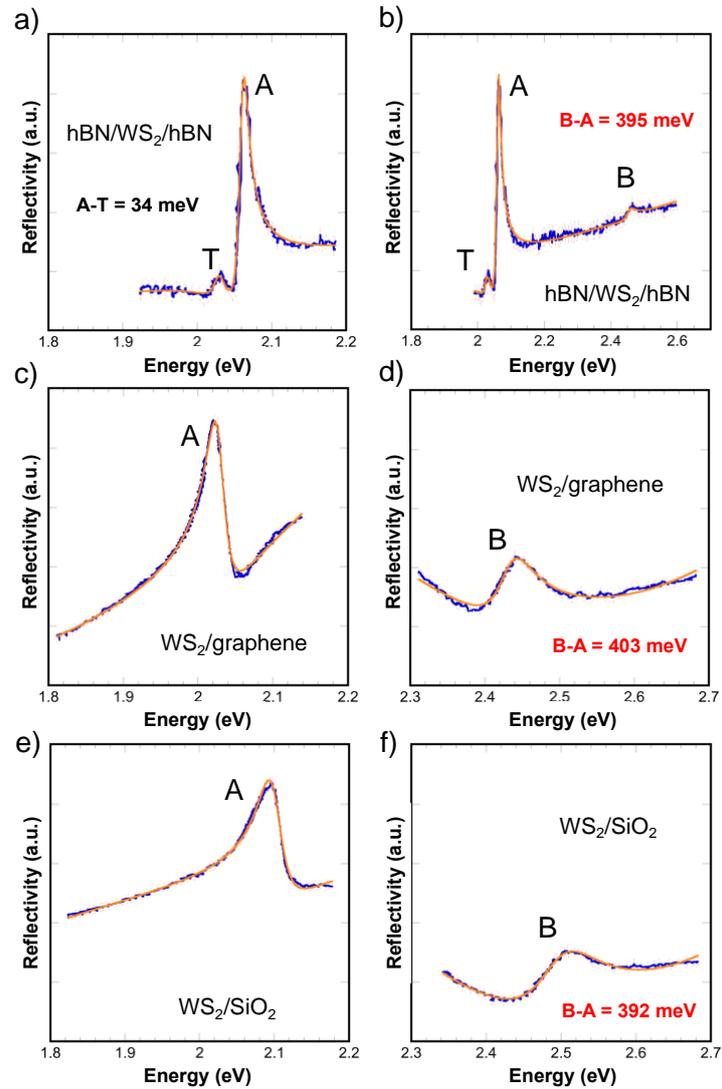

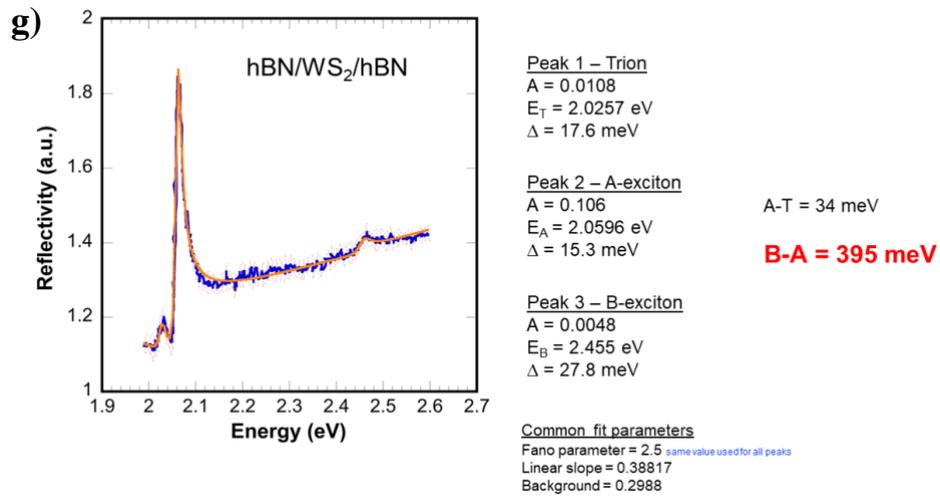

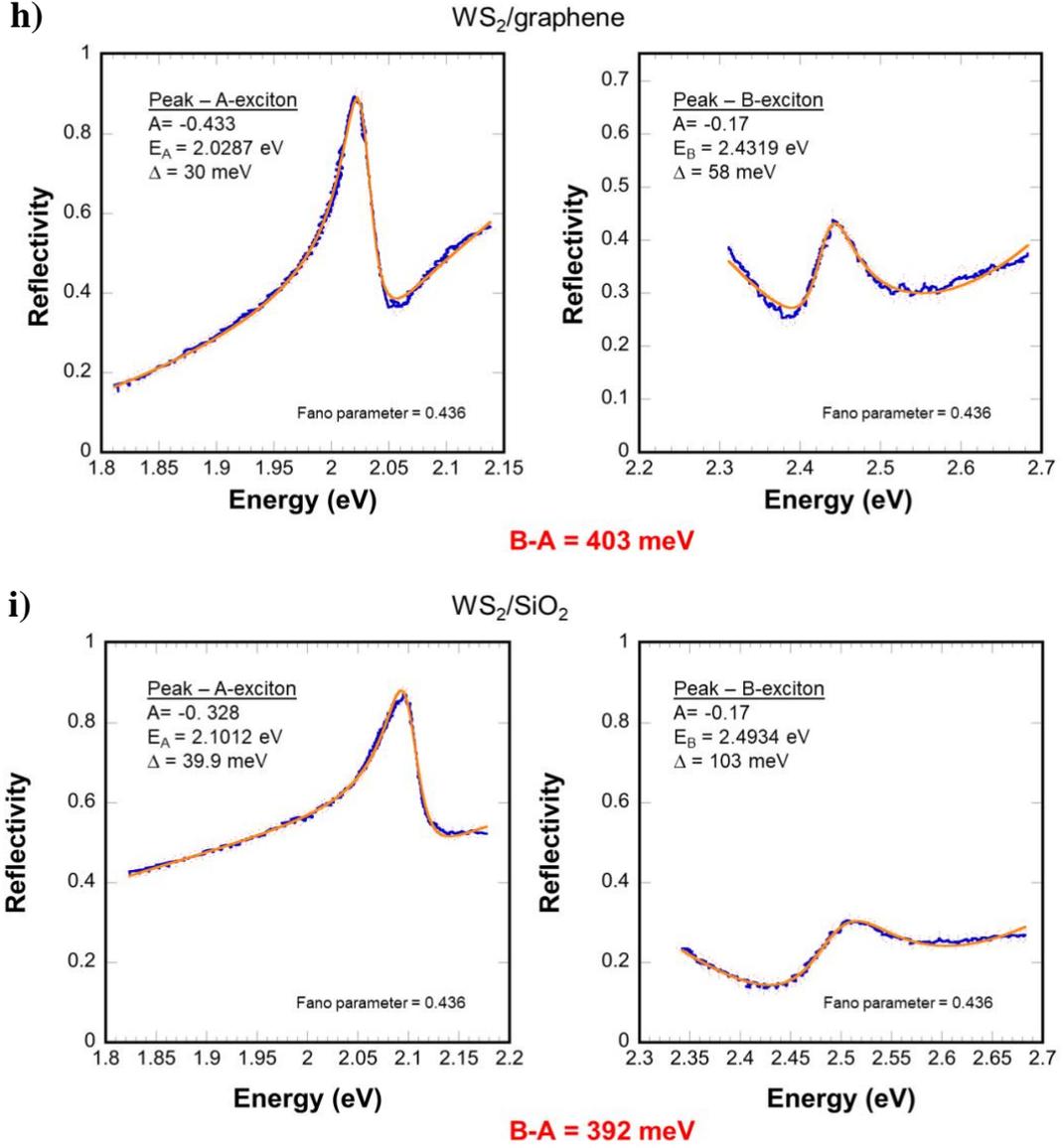

**h)**

WS$_2$/graphene

Peak – A-exciton
A= -0.433
E$_A$ = 2.0287 eV
Δ = 30 meV

Fano parameter = 0.436

Peak – B-exciton
A= -0.17
E$_B$ = 2.4319 eV
Δ = 58 meV

Fano parameter = 0.436

**B-A = 403 meV**

**i)**

WS$_2$/SiO$_2$

Peak – A-exciton
A= -0.328
E$_A$ = 2.1012 eV
Δ = 39.9 meV

Fano parameter = 0.436

Peak – B-exciton
A= -0.17
E$_B$ = 2.4934 eV
Δ = 103 meV

Fano parameter = 0.436

**B-A = 392 meV**

**Fig. S1:** Differential reflectivity analysis of hBN/WS$_2$/hBN (a,b), WS$_2$/Gr (c,d) and WS$_2$/SiO$_2$ (e,f). The energy difference between A and B exciton is indicated in each case. The fitting equation for the A and B exciton is $f(x) = A \frac{\left(q\frac{m}{2}+x-\mu\right)^2}{\left(\frac{m}{2}\right)^2+(x-\mu)^2} + kx + b$, where $A$ is the amplitude, $q$ is the Fano parameter which represents the ratio of resonant scattering to the background scattering, $m$ is the width of the line shape, $k$ is the linear slope and $b$ is the intercept.[1] The corresponding energies and linewidths of A and B excitons are presented after fitting (orange

lines) of the experimental data (blue lines), for the case of g) hBN/WS$_2$/hBN, h) WS$_2$/graphene and i) WS$_2$/SiO$_2$.

## *B. WS$_2$/graphene – a 6-fold suppression of the X$^0$ PL intensity*

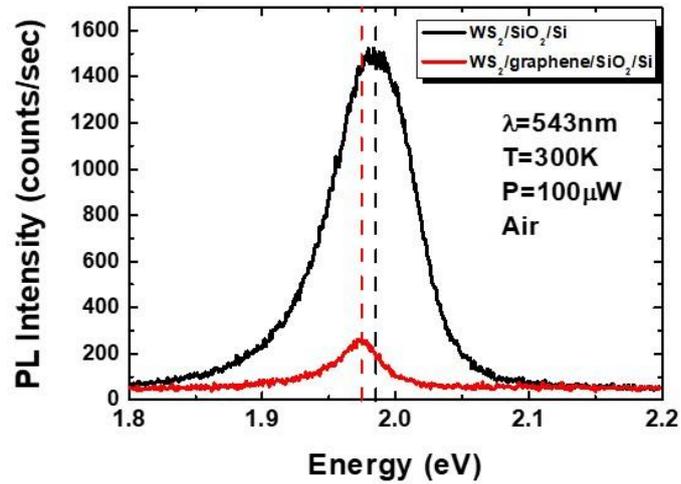

**Fig.S2:** Comparison of the PL emission between WS$_2$/SiO$_2$ and WS$_2$/graphene under 543nm excitation at 300K. A 6-fold decrease in the case of WS$_2$/graphene is observed.

## *C. PL analysis*

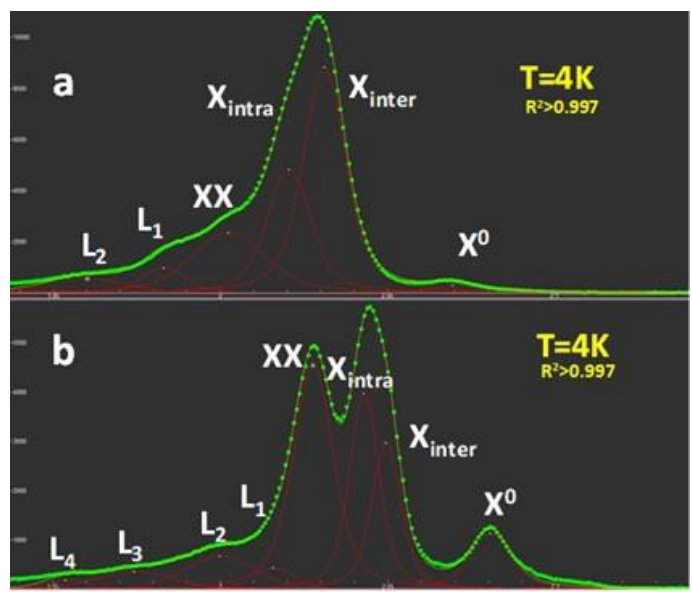

**Fig.S3 (a) , (b):** Voigt fitting of the PL emission at 4K under 543nm excitation in the cases of a) hBN/WS$_2$/hBN and b) WS$_2$/hBN.

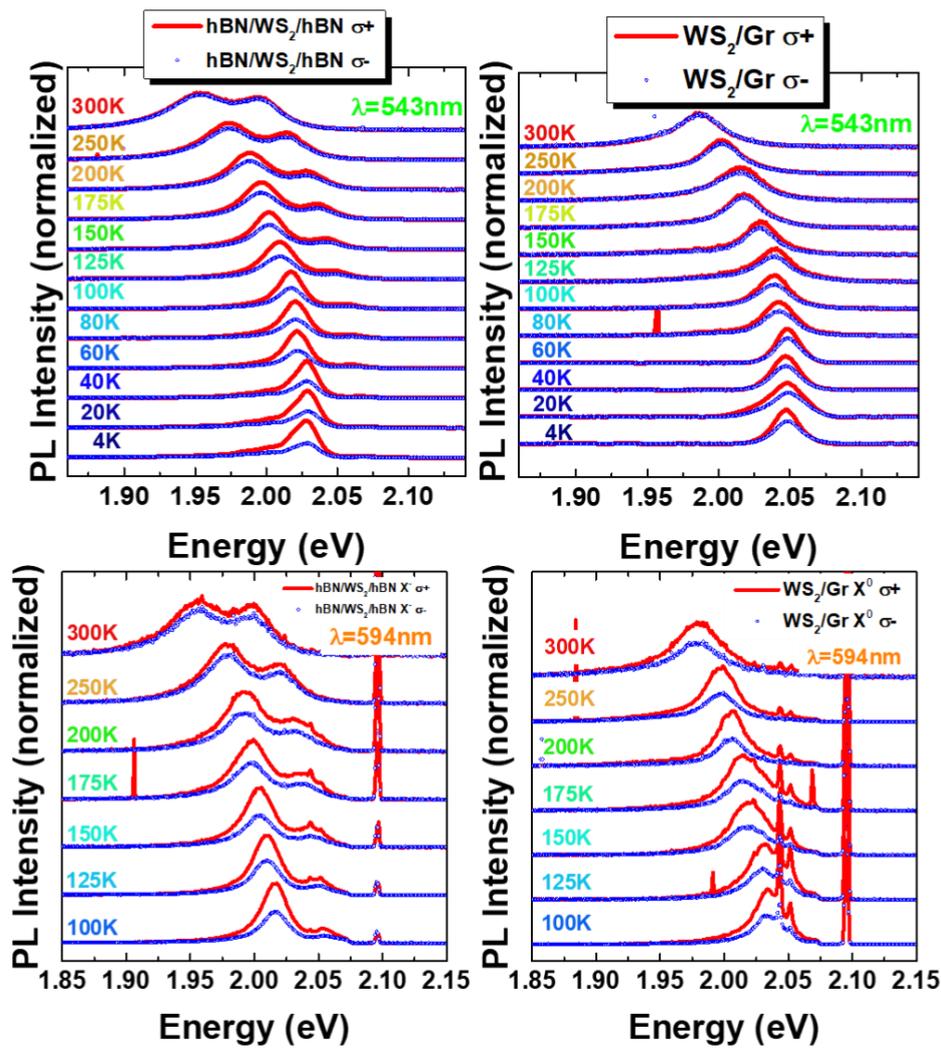

**Fig.S3(c):** Polarization-resolved PL spectra of hBN/WS$_2$/hBN and WS$_2$/Gr under non-resonant (top panels - 543nm) and near-resonant (bottom panels - 594nm) excitation.

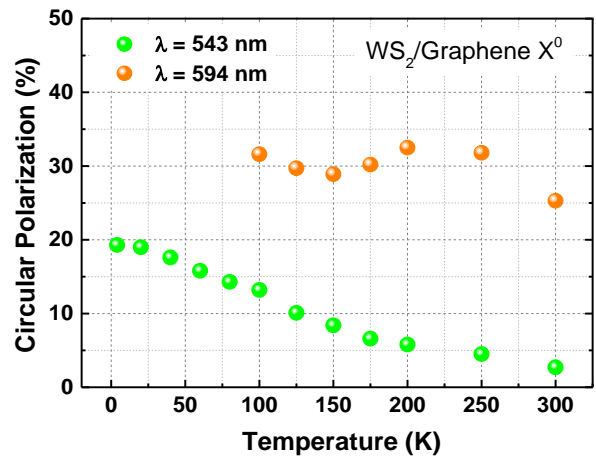

**Fig.S3(d):** T-dependent valley-polarization for X$^0$ of WS$_2$/Graphene under 543nm and 594nm excitation extracted from a second sample.

## D. Raman measurements

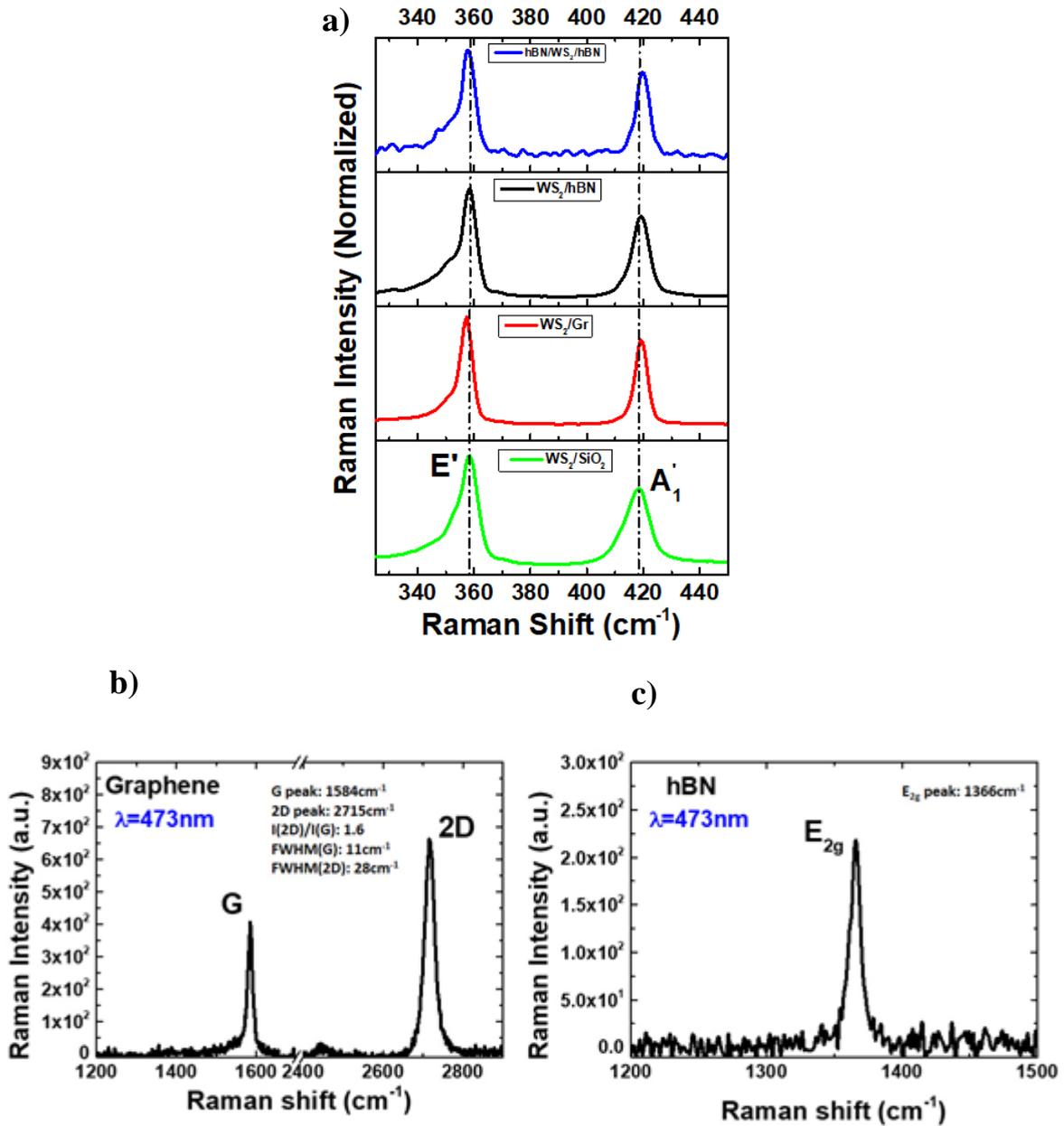

**Figure S4:** Raman spectra of (a) WS$_2$/SiO$_2$, WS$_2$/Gr, WS$_2$/hBN, and hBN/WS$_2$/hBN structures, b) CVD graphene and c) hBN acquired with a 473nm laser.

Strain is a parameter that could affect the band structure of WS$_2$, therefore we utilized Raman spectroscopy to characterize WS$_2$ to identify possible strain effects in the different cases studied. A possible source of strain is the lattice constant mismatch of WS$_2$ with graphene and hBN. Room temperature Raman spectra of the two main vibrational modes of $WS_2$ (*i.e.* in-plane mode E′ and out-of-plane mode $A_1$′) are presented in Fig. S4a. Compared to the $SiO_2$ substrate, a systematic red shift of the E′ mode and a blue shift of the $A_1$′ mode is observed due to an increase of the weak interaction with atoms from neighbouring layers (graphene and hBN) and a raise in the effective screening of the Coulomb potential.[2] In addition, a clear decrease in the linewidth of the vibrational modes is observed on hBN and graphene compared to the $SiO_2$ substrate. Due to the larger lattice constant of $WS_2$ (3.15 Å)[3] compared to graphene (2.46 Å)[4] and hBN (2.5 Å)[5], we would expect that contraction should occur, accompanied with a blue shift of the E′ mode. However, the ~1.7 cm$^{-1}$ red shift of the E′ in the $WS_2$/graphene and hBN/$WS_2$/hBN samples suggests that the effective screening of the Coulomb potential dominates any effect due to lattice mismatch. Such an effect was seen in $MoS_2$ and attributed specifically to the dielectric screening of the long-range components of the Coulomb forces. In Fig. S4b and Fig.S4c we present Raman spectra from the underlying CVD graphene and hBN respectively. The position and the linewidth of the G and 2D peaks, as well as the intensity ratio ($I_{2D}/I_G$) confirm a p-type graphene in our heterostructures.[6]

### E. Raman contribution to PL spectra

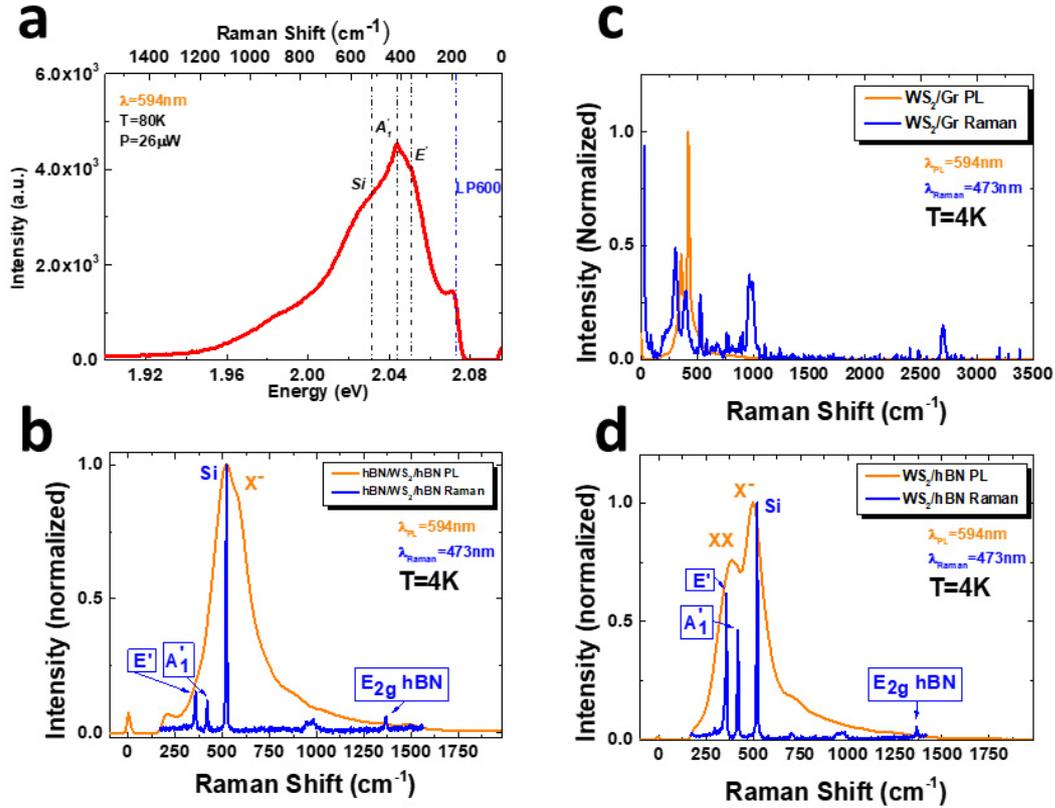

**Figure S5:** Raman contribution in the PL emission at low temperatures under resonant excitation in the cases of a) WS$_2$/SiO$_2$, b) hBN/WS$_2$/hBN, c) WS$_2$/graphene and d) WS$_2$/hBN.

Helicity resolved PL spectra measured below 100 K under 594nm excitation (Fig.2c and 2d in the main text) are not included in our analysis because the contribution of the active Raman modes in the photoluminescence spectra and their strong effect in the $\sigma_+$ and $\sigma_-$ emission make the extraction of the degree of polarization unreliable. In Fig. S5a we present photoluminescence spectra of hBN/WS$_2$/hBN and WS$_2$/Graphene excited with a 594 nm laser at 4 K. In the same plot we have included Raman spectra acquired with a 473 nm laser to demonstrate the pure Raman signal. It is apparent that the strong underlying vibrational modes have a significant contribution in the PL emission when the excitation is near-resonant. In addition, the cross-polarized helicity of the Raman modes (due to mirror backscattering effects) affect strongly the PL polarization analysis. Therefore, we only present the degree of polarization above 100 K.

## F. Screening effects due to carrier doping

To model the monotonic temperature dependence of the degree of polarization, we consider screening effects due to carrier doping. WS$_2$ is in general an n-type semiconductor and we confirm this character in our monolayers by extracting the binding energy of the X$^-$ on WS$_2$/SiO$_2$ (~38meV). Theoretical calculations[7] have considered the effect of doping on the valley relaxation by using a statically screened Coulomb potential with a finite Thomas-Fermi wave vector ($k_{TF}$), which is related to the carrier density $n$ by

$$k_{TF} \equiv k_{TF}^0 \left[ 1 - e^{-2\pi\hbar^2 n/(g_s g_v m^* k_B T)} \right] \qquad (S1), [7]$$

where $k_{TF}^0 \equiv g_s g_v m^* e^2/(4\pi\varepsilon\hbar^2)$ is the zero-temperature Thomas-Fermi wave vector, $g_s$ ($g_v$) is the degeneracy number for spins (valleys), $m^*$ is the carrier effective mass, and $\varepsilon$ is the dielectric constant which is different for the 4 cases studied in this work.

Under non-resonant excitation conditions, the temperature dependence of the polarization degree (Fig. 1d) was fitted using equation S3, derived by considering the steady state of a rate equation model. Both characteristic times, the exciton total lifetime $\tau_r$ and the spin-valley relaxation time $\tau_v$, depend on temperature. If the former is dominated by the radiative decay, then it is just proportional to the temperature, if there are no dark states. Otherwise, radiative decay shows a non-monotonic behavior in the presence of dark states. The latter, $\tau_v$, has a nontrivial dependence through the Thomas-Fermi wave vector and the homogeneous broadening[7]

$$\tau_v \sim k_{TF}^2 \tau_r \qquad (S2)$$

According to Eq. (1) of the main text the polarization rate should follow:

$$P_{circ} = \frac{P_0}{1 + \dfrac{A}{\left(1 - e^{\frac{-0.8 \cdot n}{T}}\right)^2}} \qquad (S3)$$

where A is a fit parameter fixed at 0.0005. We also fixed the carrier density at $4 \times 10^{12} cm^{-2}$ in the two cases and varied the low temperature polarization $P_0 \approx P(4\text{K})$. In the hBN/WS$_2$/hBN sample, we used $P(4K) = 40.84$ whereas in the WS$_2$/Gr, $P(4K) = 18.62$. For carrier densities of the order of $10^{12} cm^{-2}$ an almost linear dependence of the intervalley scattering time as a function of temperature is observed and our experimental results are well fitted with this model (Fig.F1). Nevertheless, we should comment that taking into account similar doping densities in WS$_2$ when it is encapsulated in hBN or on top of graphene is not a realistic scenario since graphene certainly affects the electron density of WS$_2$. On the other hand, if we change the electron density $n$ in the Thomas-Fermi wave vector equation (S1), we do not get a reliable fit. A reason for this could be the competition of the screening due to electrons with other effects, such as the dielectric screening from hBN or graphene, the activation of dark-to-bright exciton path at elevated temperatures, the electron-phonon coupling and the energy difference between the excitation and emission which affects the relaxation mechanisms. For instance, it is obvious that the temperature dependent spin-valley polarization under near-resonant excitation shown in Fig. 1e cannot be fitted with the same equation, i.e. taking into consideration only screening effects that will affect the valley lifetime. Therefore, this becomes a complicated problem and a simple model cannot explain the observations.

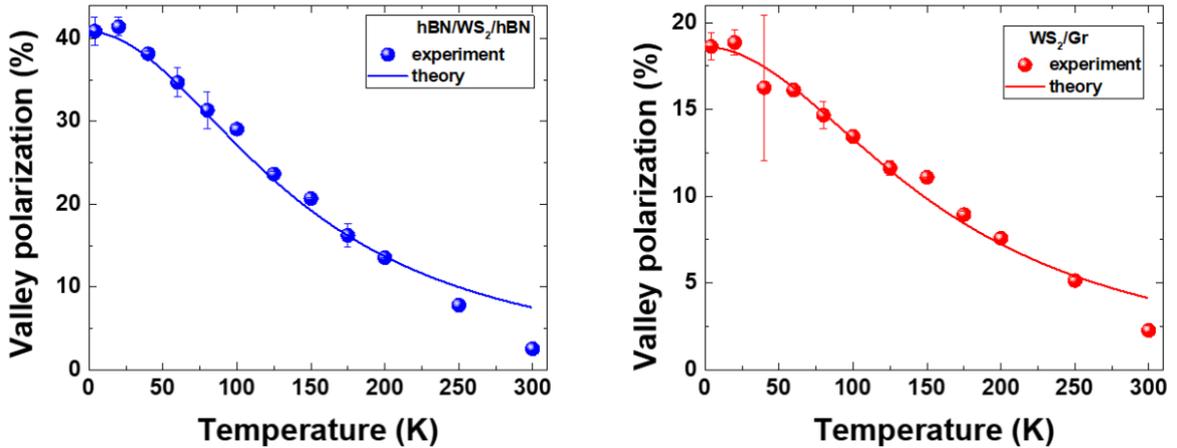

### G. The band-proximity effect

Band structures are obtained within the frame work of plane-wave density function theory using the open source code Quantum Espresso.[8] Fully relativistic norm-conserving pseudopotential is used to include spin-orbit coupling. The cutoff energy is set to 60 Ry and Brillouin zone samplings is approximated by 16×16×1 k-points. Crystal structures of WS$_2$ heterostructures are shown in fig. (S6).

(a)                                        (b)

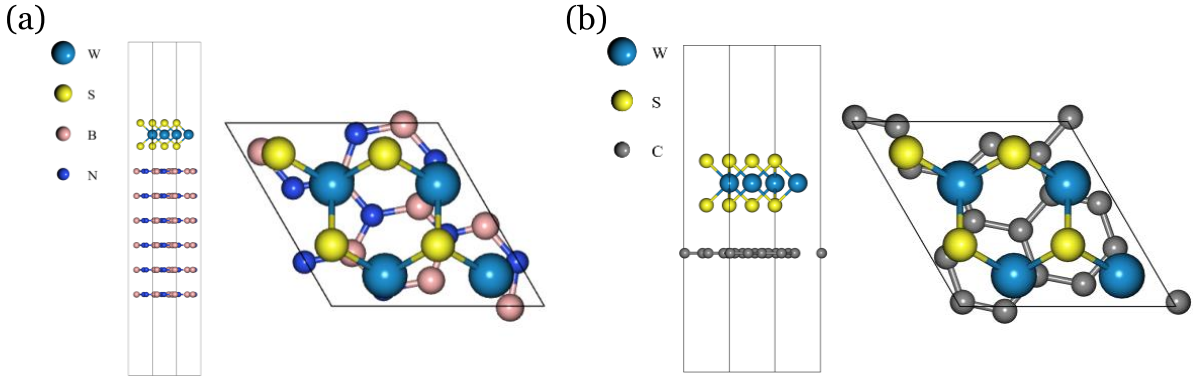

**Figure S6:** Crystal structures of (a) $2 \times 2$ WS$_2$ on $\sqrt{7} \times \sqrt{7}$ hBN and (b) $2 \times 2$ WS$_2$ on $\sqrt{7} \times \sqrt{7}$ graphene. Lattice constants of WS$_2$, hBN and graphene of 3.18 Å, 2.5 Å, and 2.46 Å, respectively, correspond to the strains in hBN and graphene of $-4.14\%$ and $-2.35\%$, correspondingly.

### H. Phonon limited scattering rate

The schematic structure of heterostructures is depicted in Fig. S7. Atomically thin materials 1 and 2 are bounded between two semi-infinite substrate materials 3 and 4, characterized by their dielectric function. The thickness of materials 1 and 2 are $t_1$ and $t_2$, respectively. $\epsilon_n$ the dielectric functions of the mediums $n$, where $n = 1,2,3,4$:

$$\epsilon_n(\omega) = \epsilon_n(\infty)\frac{\omega^2 - \omega_{LO,n}^2}{\omega^2 - \omega_{TO,n}^2}. \qquad (S4)$$

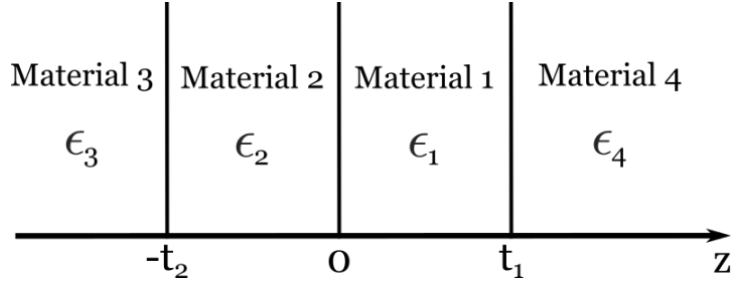

**Figure S7:** Schematic structure of heterostrutures.

We look for the solution of Maxwell equations in the form of[9]

$$\phi(\boldsymbol{\rho}, z) = \sum_{\boldsymbol{q}} \phi(q,z) e^{i\boldsymbol{q}\cdot\boldsymbol{\rho}}, \qquad (S5)$$

Where

$$\phi(q,z) = \begin{cases} Ae^{-q(z-t_1)} & t_1 < z \\ Be^{q(z-t_1)} + Ce^{-qz} & 0 < z \le t_1 \\ De^{qz} + Ee^{-q(z+t_2)} & -t_2 < z \le 0 \\ Fe^{q(z+t_2)} & z \le -t_2 \end{cases} \qquad (S6)$$

Continuity of the potentials, $\phi_n$, and displacement fields, $\epsilon_n \frac{\partial \phi_n(z)}{\partial z}$ at the boundaries are used to express coefficients $A$, $B$, $D$ and $F$ in eq. (S6) in term of $C = \phi_0$ and the frequencies from the secular equation:

$$\frac{(\epsilon_1 + \epsilon_4)}{[(\epsilon_1 + \epsilon_2)(\epsilon_2 - \epsilon_3)\exp(-2qt_2) + (\epsilon_1 - \epsilon_2)(\epsilon_2 + \epsilon_3)]\exp(-qt_1)} + \frac{(\epsilon_4 - \epsilon_1)\exp(-qt_1)}{(\epsilon_2 - \epsilon_3)(\epsilon_1 - \epsilon_2)\exp(-2qt_2) + (\epsilon_1 + \epsilon_2)(\epsilon_2 + \epsilon_3)} = 0$$
$$(S7)$$

$$\phi(q,z) = \phi_0 \begin{cases} (\alpha_1 + 1)e^{-qz} & t_1 < z \\ \alpha_1 e^{q(z-2t_1)} + e^{-qz} & 0 < z \le t_1 \\ (1 + \alpha_1 e^{-2qt_1})[(1 - \alpha_2 e^{-qt_2})e^{qz} + \alpha_2 e^{-q(z+t_2)}] & -t_2 < z \le 0 \\ (\alpha_2(1 - e^{-2qt_2}) + e^{-qt_2})(1 + \alpha_1 e^{-2qt_1})e^{q(z+t_2)} & z \le -t_2 \end{cases} \qquad (S8)$$

Where $\alpha_1 = \frac{\epsilon_1 - \epsilon_4}{\epsilon_1 + \epsilon_4}$ and $\alpha_2 = \frac{(\epsilon_2 - \epsilon_3)\exp(-qt_2)}{\epsilon_2 + \epsilon_3 + (\epsilon_2 - \epsilon_3)\exp(-2qt_2)}$.

$\phi_0$ can be found by relating energy of the excitation quanta $\hbar\omega_q$ to the electromagnetic field energy[10]

$$\phi_0 = \sqrt{\frac{4\pi\hbar}{qA}} \left[ \frac{\partial\epsilon_1}{\partial\omega}(1 - e^{-2qt_1})(1 + \alpha_1^2 e^{-2qt_1}) + \frac{\partial\epsilon_2}{\partial\omega}(1 + \alpha_1 e^{-2qt_1})^2(1 - e^{-2qt_2})((1 - \alpha_2 e^{-qt_2})^2 + \alpha_2^2) + \frac{\partial\epsilon_3}{\partial\omega}(\alpha_2(1 - e^{-2qt_2}) + e^{-qt_2})^2(1 + \alpha_1 e^{-2qt_1})^2 + \frac{\partial\epsilon_4}{\partial\omega}(\alpha_1 + 1)^2 e^{-2qt_1} \right]^{-1/2}$$
$$\text{(S9)}$$

Where $A$, area of the sample and $\hbar$ is reduces plank constant. The electron-phonon Hamiltonian can be written as[10]

$$H = \sum_q -e\phi(q,z)e^{iq.\rho}\left(a_q + a_{-q}^\dagger\right) \qquad \text{S(10)}$$

Where $-e\phi(q,z)$ is known as an electron phonon coupling factor. $a_q$ $\left(a_{-q}^\dagger\right)$ is annihilation (creation) operator and $e$ is the elementary charge. Scattering rate is calculated using Golden rule:

$$\frac{1}{\tau_{k(T)}} = \frac{2\pi}{\hbar} \sum_{k'} |\Gamma(q,z)|^2 \times \left[ N(T)\delta(\varepsilon_{k'} - \varepsilon_k - \hbar\omega) + (N(T) + 1)\delta(\varepsilon_{k'} - \varepsilon_k + \hbar\omega_q) \right]$$
$$\text{S(11)}$$

Where $q = |\mathbf{q}| = |\mathbf{k} - \mathbf{k'}|$ is a momentum transfer. The energy dispersion $\varepsilon_k$ is approximated using parabola $\varepsilon_k = \frac{\hbar^2 k^2}{2m}$ with an effective mass $m = 0.34\, m_e$. $N(T)$ is Bose-Einstein distribution function and $\omega_q$ is a solution of eq. (S7). By considering a homogenous electron distribution in the atomically thin $WS_2$, $\Gamma(q,z)$ can be written as

$$\Gamma(q,z) = -e\phi_0 \frac{1}{t_1} \int_0^{t_1} \left( \alpha_1 e^{q(z - 2t_1)} + e^{-qz} \right) dz \qquad \text{S(12)}$$

Dielectric function of graphene in layer 2, where applicable, is calculated using random-phase approximation

$$\epsilon_{Gr}(q,T) = 1 + v_c \Pi(q,T), \ ^{11} \qquad (S13)$$

where $v_c = 2\pi e^2/Kq$, $K$ is environment lattice dielectric constant. $\Pi(q,T)$ is the polarization function.[11]

The inter-valley scattering rates of electrons in $WS_2$ hetero-structures are shown in Fig. S8, where the thickness of $WS_2$ is assumed to be 6.14 Å.[12] The Brillouin zone sampling for electronic states is $200 \times 200$ k-points. The corresponding parameters for the dielectric functions are listed in table S1.[12–14]

| | $\hbar\omega_{TO}\ [meV]$ | $\hbar\omega_{LO}\ [meV]$ | $\epsilon(0)$ | $\epsilon(\infty)$ |
|---|---|---|---|---|
| $WS_2$ | 42.9 | 43 | 13.7 | 13.6 |
| hBN | 97.6 | 103 | 5.1 | 4.58 |
| $SiO_2$ | 55.7 | 60 | 3.9 | 3.36 |

**Table S1:** Energies and dielectric constants of used materials.

Fig. S7a shows scattering rate from the bottom of the conduction band at the K point to all available states near the K' point. As the temperature increases more phonons become available and the scattering rate grows as N(T) as shown in Fig. S7a. For intermediate initial state energies enough to excite optical phonon in $WS_2$ or $SiO_2$, scattering rate is finite even at the very low energies and it keeps increasing as more phonons become thermally populated as shown in Fig. S7b. This situation is applicable to the near resonant excitation conditions with the laser wavelength of 594 nm used in the experiment. As the excitation energy becomes large enough to scatter an optical phonon hBN at low temperatures the phonon inter-valley decay rate becomes much larger in hBN encapsulated sample due to the stronger electron-phonon scattering. This situation corresponds to the non-resonant excitation conditions with excitation laser wavelength of 543 nm and it is shown in Fig. S7c. Note that the exciton scattering rates are about twice as large since both an electron and a hole forming an exciton can scatter a phonon.

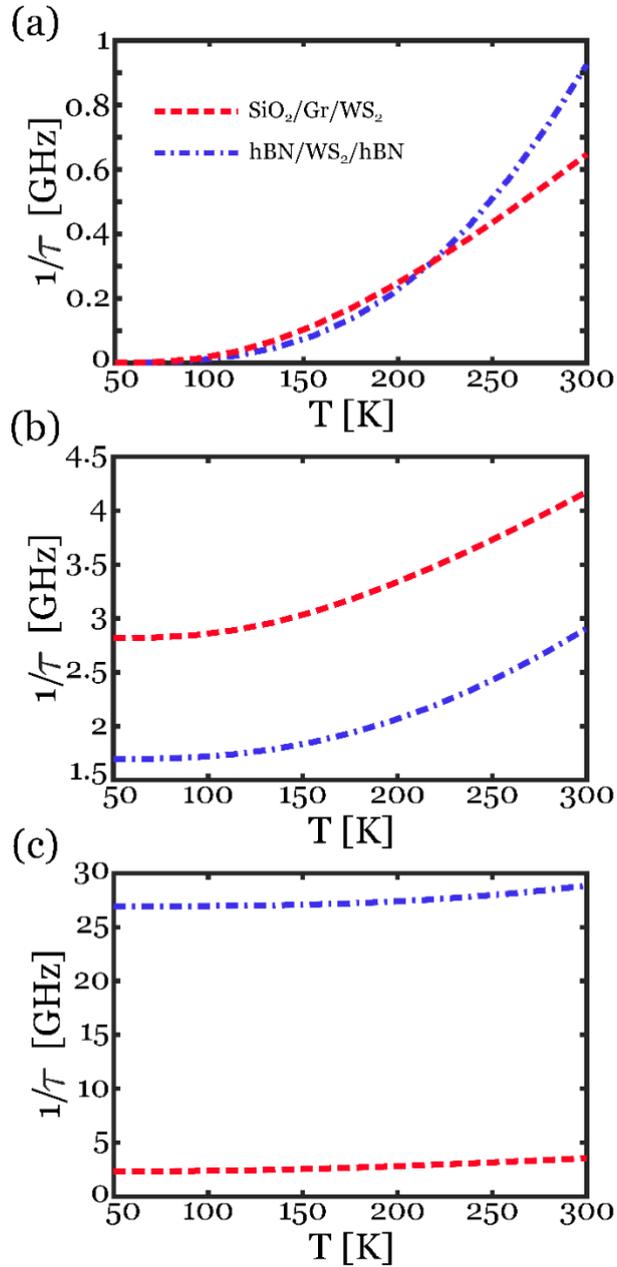

**Figure S7:** Inter-valley scattering rates of electrons in $WS_2$ heterostructures. (a) $\varepsilon_k = 0\ meV$, (b) $\varepsilon_k = 55\ meV$, and (c) $\varepsilon_k = 200\ meV$.

The bandgap energy shift due to electron-phonon coupling is calculated from the real part of the conduction band electron self-energy, similar to Ref.[15] The results are shown in Fig. S8, where the temperature dependence of the bandgap shift is much weaker than observed in the experiment. This suggests that other than just Fröhlich coupling to phonons and interband scattering (virtual valence to conduction transitions[16]) are responsible for the observed bandgap renormalization.

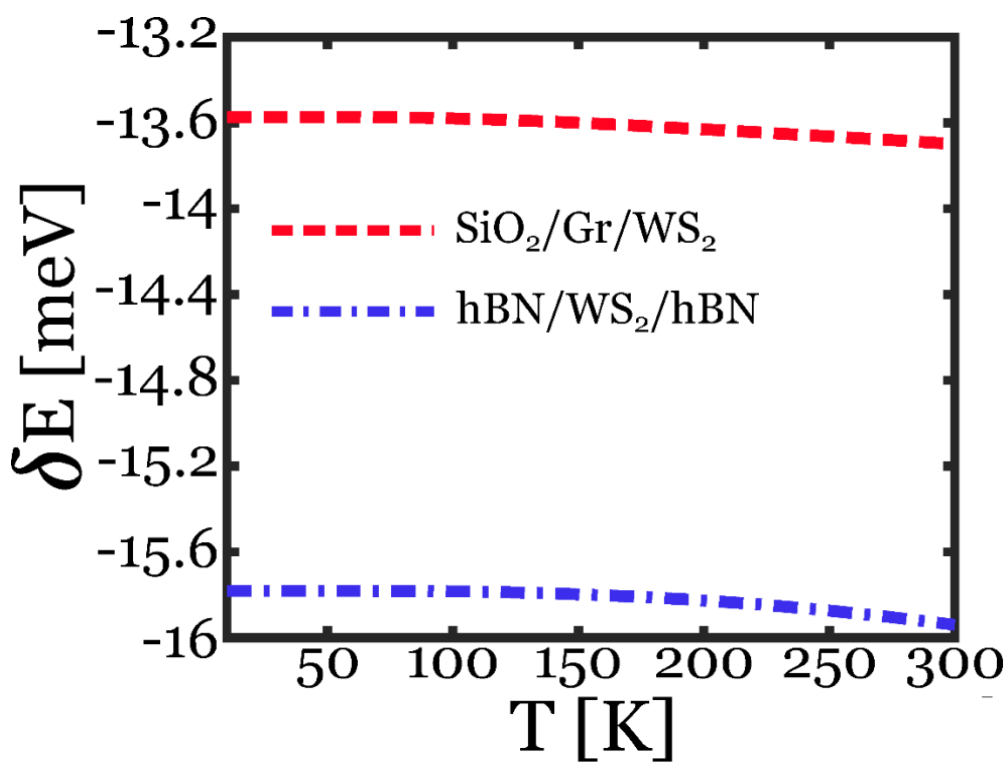

**Figure S8:** Calculation of the T-dependent band-gap renormalization of $SiO_2$/Gr/$WS_2$ and hBN/$WS_2$/hBN due to the Fröhlich coupling to phonons.